\documentclass[a4paper]{scrartcl}
\usepackage[T1]{fontenc}
\usepackage[utf8]{inputenc}
\DeclareUnicodeCharacter{200B}{{\hskip 0pt}}
\DeclareUnicodeCharacter{2082}{\textsubscript{2}}
\usepackage[noblocks,auth-sc]{authblk}
\usepackage{url}
\usepackage[breaklinks,hidelinks,linktoc=all]{hyperref}
\usepackage{graphicx}
\usepackage{cleveref}
\usepackage{xcolor}
\usepackage[textsize=tiny,tickmarkheight=1em]{todonotes}

\makeatletter
\renewcommand{\AB@affilsepx}{\protect,\ \protect\Affilfont}
\makeatother

\newcommand{\arxiv}[1]{\href{https://arxiv.org/abs/#1}{\nolinkurl{arXiv:#1}}}

\title{Shaping the Digital Future of ErUM Research: Sustainability \& Ethics}
\subtitle{Workshop Report}
\date{October 2025}

\author[1,2]{Luca Di Bella}
\author[1]{Jan Bürger}
\author[3]{Markus Demleitner}
\author[4,5]{Torsten Enßlin}
\author[6]{Johannes Erdmann}
\author[1,6]{Martin Erdmann}
\author[1,6]{Benjamin Fischer}
\author[7]{Martin Gasthuber}
\author[6]{Gabriele Gramelsberger}
\author[​]{Wolfgang Gründinger}
\author[8]{Prateek Gupta}
\author[​]{Johannes Hartl}
\author[9]{Maximilian Horzela}
\author[7]{Vijay Kartik}
\author[6]{Stefan Krischer}
\author[10]{Eva Kröll}
\author[11]{Thomas Kuhr}
\author[2]{Katharina Kürschner}
\author[9]{Inga Lakomiec}
\author[12]{Valerie Lang}
\author[13]{Kristin Lohwasser}
\author[14]{Thomas Metcalf}
\author[15]{Martin Möller}
\author[6]{Saskia Nagel}
\author[16]{Susanne Pfalzner}
\author[10]{Rebecca Redlin}
\author[​]{Christopher Schrader}
\author[7]{Kathrin Schulz}
\author[12]{Markus Schumacher}
\author[7]{Kilian Schwarz}
\author[4]{Fabian Sigler}
\author[7]{Dwayne Spiteri}
\author[6]{Achim Stahl}
\author[1]{Judith Steinfeld}
\author[16]{Wim Vanderbauwhede}
\author[2]{Cyrus Walther}
\author[1]{Angela Warkentin}
\author[6]{Peter Wissmann}
\author[17]{Eoin Woods}

\affil[1]{ErUM-Data~Hub}
\affil[2]{TU~Dortmund~University}
\affil[3]{Heidelberg~University}
\affil[4]{Max~Planck~Institute~for~Astrophysics}
\affil[5]{German~Center~for~Astrophysics}
\affil[6]{RWTH~Aachen~University}
\affil[7]{DESY}
\affil[8]{Thuringian~State~Observatory}
\affil[9]{Georg-August-Universität~Göttingen}
\affil[10]{PT.DESY}
\affil[11]{LMU~Munich}
\affil[12]{University~of~Freiburg}
\affil[13]{University~of~Sheffield}
\affil[14]{University~of~Bonn}
\affil[15]{Öko-Institut~Consult~GmbH}
\affil[16]{University~of~Glasgow}
\affil[17]{Artechra~Sdn~Bhd}

\begin{document}
\maketitle
\pagebreak
\section*{Abstract}
\begin{abstract}
This workshop report from “Shaping the Digital Future of ErUM Research: Sustainability \& Ethics” (Aachen, 2025) reviews progress on sustainability measures in data-intensive ErUM-Data research since the 2023 call-to-action on resource-aware research. It evaluates short-, medium-, and long-term actions around monitoring and reducing CO₂ emissions, improving data and software FAIRness, optimizing workflows and computing infrastructures, and aligning operations with low-carbon energy availability, including concepts such as “breathing” computing centers, long-term data storage strategies, and software efficiency certification. The report stresses the need for systematic teaching, training, mentoring, and new support formats to establish sustainable coding and computing practices, particularly among students and early-career researchers, and highlights the importance of dedicated steering and funding instruments to embed sustainability in project planning. Ethical discussions focus on the transformative use of AI in ErUM-Data, addressing autonomy, bias, transparency, explainability, attribution of responsibility, and the risk of deskilling, while reaffirming that accountability for scientific outcomes remains with human researchers. Finally, the report emphasizes that sustainable transformation requires not only technical measures but also targeted awareness-building, communication strategies, incentives, and community-driven initiatives to move from awareness to action and to integrate sustainability and ethics into everyday scientific practice.    
\end{abstract}
\pagebreak
\tableofcontents
\pagebreak

\section{Introduction}\label{introduction}

Digital technologies have always transformed the way science is conducted and continue to shape scientific practice.
Two topics stand out as currently dominating the conversation as challenging and in need of deliberation: Sustainability and Artificial Intelligence.
They were at the center of a workshop held in Aachen from July 28th to August 1st, 2025~\cite{sustainability_workshop2025} for scientists working on digitization in the research on universe and matter in Germany (Erforschung von Universum und Materie - ErUM).
This workshop followed an earlier one addressing the global climate challenge~\cite{sustainability_workshop2023}, resulting in the publication “Resource-aware research on Universe and Matter: call-to-action in digital transformation”~\cite{Bruers2024}.
This workshop report details the outcome of the discussion around the challenges and opportunities surrounding Sustainability and Artificial Intelligence in digital research.
In the following, the progress since the first workshop is reviewed and new developments are discussed subsequently.

\subsection{Progress Made}\label{progress-made}

In the previous publication~\cite{Bruers2024}, measures were categorized in three time scales on which they could be implemented, short (S), medium (M), and long (L).
Here we list all measures and to what extent they were implemented.

Measures that were considered to be implementable immediately or on short timescale with little effort are:

\begin{itemize}
\item \textbf{S1: Raise awareness of the climate challenge at all levels.}

Awareness has increased due to initiatives of community members and led to a higher visibility of the topic.
For example, sustainability is discussed in the input of the German particle physics community to the Update of the European Strategy of Particle Physics~\cite{ESPPU_input}.
The European Astronomical Society Meeting included a session dedicated to sustainability~\cite{EAS_sustainability_session}.
Following the in-person workshop in 2023, the ErUM-Data-Hub organized two online discussion afternoons to facilitate exchange on sustainability-related topics and initiatives ~\cite{edh-sus-2024,edh-sus-2024-2}.
Another example is the \textit{Know your footprint} tool~\cite{Lang:2024wpo} that allows scientists to estimate the footprint of their research.
The ATLAS collaboration has produced, for the first time at the LHC, a paper reviewing the sustainability of ATLAS computing~\cite{ATLAS:2025sgg}.
In the Belle II collaboration, a sustainability task force was created to quantify the footprint of the experiment and to suggest measures for its reduction.

\item \textbf{S2: Disseminate knowledge of measures to address the challenge.}

The authors of the first report~\cite{Bruers2024} received invitation to various conferences where they disseminated ideas on how to address sustainability in computing discussed at the original workshop.
These ideas were picked up and brought forward grant proposals, for example during the recent ERUM-Data funding round.

\item \textbf{S3: Monitor and report energy consumption at job level.}

The monitoring and reporting of the per job energy consumption is becoming more widespread at sites.
For example, DESY provides an estimate of the CO$_2$ footprint of batch jobs to their users.
For the VISPA cluster of the ErUM-Data groups in Aachen a system was developed, that not only informs users about the footprint of their jobs, but also gives indications when the energy mix is most favorable for running jobs.

At experiment level, ATLAS is a good example.
Its workflow management system, PanDA, has been equipped to collect estimates of the footprint of all ATLAS jobs.
Users are then provided an estimate of the footprint of their job, along with a link to more information about the calculations~\cite{ATLAS:2025sgg}.
While the actual footprint depends on the site where the job runs, the numbers reported to users are based on averages over all sites to discourage people from sending jobs to low emission sites and clogging those up. 
This way a more stable estimate can be provided so that users can better assess the impact that their analysis code and workflow has.

\item \textbf{S4: Consider carbon footprint for all investments and project plans.}

Footprint considerations play an increasing role in the planning of new, large scale projects.
For example, as reported at the workshop, sustainability was an important factor in the planning of the computing centre for the Einstein Telescope.
For ongoing large scale projects, existing boundary conditions often make it harder to take sustainability aspects into account.
On smaller scales it usually depends on the involved persons.
Also scientific competition and financial constraints can outweigh sustainability considerations.

\item \textbf{S5: Enhance awareness of the trade-off between research benefit and climate impact.}

This turned out to be more difficult in the education of young researchers.
Teaching focuses initially on the science topic.
The young researchers first have to develop the skills to do their science and only if they have sufficient experience they can judge the trade-offs with other goals. This is particularly important for bachelor's and master's students, as the initial trials with physics analyses account for a significant proportion of the computing power used. At PhD level students usually have reached a maturity to consider sustainability aspects.

\item \textbf{S6: Use already optimized algorithms in open-source standard libraries or domain-specific libraries.}

This is becoming more and more common practice.
Fortunately, in particular young researchers (starting with PhD level) actually follow this as it turns out to be much more efficient both in saving development time and in applying the algorithms.
\end{itemize}

On a medium timescale of a few years the following measures were considered
realizable:

\begin{itemize}
\item \textbf{M1: Make data FAIR to promote reuse.}

An example demonstrating the progress in this area is the new reinterpretation method described in Ref.~\cite{Gartner:2024muk}.
It made it possible to derive a new scientific result from published Belle II data~\cite{Belle-II:2025lfq}.
Also new ErUM-Data projects that started in November 2025 will contribute to making data more FAIR.

\item \textbf{M2: Reduce and compress data having the anticipated scientific value of the retained information and the resource requirements in mind.}

This requires deep domain knowledge to be able to decide what data is scientifically relevant.
It gets even harder when one has to anticipate future use cases of the data, e.g. when new methods or new scientific questions may arise.
Nevertheless, this approach is more and more explored among sustainability-conscious ErUM-Data scientists as the rising data rates increase the pressure to be more efficient.

\item \textbf{M3: Optimize the choice of storing intermediate results against re-calculating them.}

The choice of storing vs. re-calculating intermediate results is not only a question of computing resources, but also of FAIRness.
For example, ATLAS has provided generator level simulation datasets on the Open Data portal so that they can be easily used by theorists.
The sharing of such simulations is systematically addressed by the LHC working group on Monte Carlo production with a dedicated subgroup on ``Data Sharing and New Workflows''.

Workflows (see also next item) play a major role here because they make it easy to switch between the alternatives of storing or re-calculating intermediate results.

\item \textbf{M4: Optimize job orchestration and scheduling in workflows.}

A good example of improved resource efficiency by job scheduling are the analysis trains used by the ALICE experiment.
Multiple, different analysis tasks are collected and executed in one go over the whole dataset.
The trains can be started when the energy mix is favorable.

A step further goes the new ErUM-Data project SUSFECIT.
It aims at including the availability of green energy in an automated job scheduling system.

\item \textbf{M5: Use workflow management to make processing FAIR.}

Various workflow management systems with different strengths are available.
To make it easier for users to take an informed decision which system to use the PUNCH consortium has provided ``A workflow management guideline''~\cite{Schmitt:2022tdt}.
In larger collaborations there is the trend to converge on a common workflow management system which has the advantage that expertise about it is built up and preserved.

\item \textbf{M6: Make software FAIR and reliable by following good software development practices and ensuring sustainable support.}

Good software development practices are better
known and are more frequently applied in the ErUM sciences, supported
by platforms such as GitHub or GitLab. Making software FAIR is also better
understood, but here practice lags behind and supporting tools need more development.
It is now easy to make software
publicly available, however, support and long-term maintenance remain a challenge, in
particular as funding and scientific career incentives often favor new developments
over services to a larger community.

\item \textbf{M7: Design software for optimized energy consumption and provide tools to measure it.}

There are several examples of software designed for optimized energy consumption,
using more efficient languages and data formats.
With awkward-arrays~\cite{awkwardarray_docs} sizeable chunks of events are processed in parallel instead of performing event-by-event calculations.
The RNtuple format in ROOT~\cite{ROOT_NIMA_1997} provides significant efficiency gains over the previous TTree format.

Tools to measure resource consumptions exist, but their application is not yet as widespread or as straightforward as other software quality control methods, such as testing.

\item \textbf{M8: Continue research on potential of AI, in particular generative and pre-trained models, or other new technologies for efficient use of resources, but balance gain of research action against resource consumption of these developments.}

The potential for speeding up simulations with AI models has been demonstrated for many cases, for example in the ErUM-Data project KISS.
Fast AI-based calorimeter simulations are used by the ATLAS~\cite{Krause:2024avx} and LHCb experiments~\cite{Barbetti:2023bvi}.

Also pre-trained models have been shown to be more efficient than a training from scratch for some cases.
For jet-based algorithms the OnmiJet-$\alpha$ foundation model is a good starting point \cite{Birk:2024knn}.
Low-Rank-Adaption (Lora) has been tested using LLMs for classification tasks with remarkable close accuracy compared to from-scratch trained networks.

A systematic collection and reporting of resource consumptions for such developments would help to better assess the overall impact.

\item \textbf{M9: Expand detailed monitoring and documentation of energy consumption and CO\textsubscript{2}e footprint in AI training and inference.}

This is not yet achieved, but being addressed by PUNCH4NFDI and the new ErUM-Data project SUSFECIT at national level and by WLCG and EGI at international level.

\item \textbf{M10: Monitor and report energy consumption at site and project level, provide information of the individual use per scientist/ project/ publication.}

Sites are making progress on this, but there is no central place yet, that gives an overview over the energy efficiency of sites.
Giving information of the use per scientist/project/publication is explored in the small-scale VISPA computing cluster.
A monitoring and reporting at project level usually still requires a tedious collection of information from various sources.
Moreover this information is not provided in a common format and may be incomplete.

\item \textbf{M11: Extend monitoring of resources beyond CO\textsubscript{2}e (water, material etc.).}

So far the focus was more on the carbon footprint as this appears to be the most pressing problem.
But plans are on the way to extend the monitoring, e.g. in WLCG, EGI, and the ErUM-Data project SUSFECIT.

\item \textbf{M12: Train scientists in good practices.}

This is addressed for example by schools organized by the ErUM-Data Hub.

It is important that not only those who are supposed to apply the good practices are aware of it, but also their supervisors must support and encourage their application.
There is still potential to train good practices more widespread and systematically, e.g. in curricula at universities.

\item \textbf{M13: Strive to become a role model at all levels and help to establish sustainability in everyday life.}

This is a highly demanding recommendation.
While progress has been made as reported by participants of the workshop, there is still room for further improvement. Progress can be seen particularly in areas where results can be measured and put into practice, e.g., in the reduction of computing power. In areas that are less quantifiable, such as communication, there is a lack of progress tracking and, as a result, a lack of incentives and motivation. Communication strategies should be better adjusted to the target group and their values.

\item \textbf{M14: Regularly review and update the CO\textsubscript{2}e reduction plan.}

This is not yet done systematically.

\end{itemize}

A longer term coordinated planning was anticipated for the following measures:

\begin{itemize}

\item \textbf{L1: Adjust computing in space and time according to the availability of renewable energy, e.g., computing centers close to off-shore wind parks with a job scheduling using only or mainly the surplus of renewable energy available at given times.}

This is a very difficult to reach goal.
We see at various centers that there are extremely well established infrastructures and excellent staff.
The idea of a new center close to off-shore has not yet gained much support.
A first approach in the direction of green job scheduling will be done by the ErUM-Data project SUSFECIT.

\item \textbf{L2: Develop software and middleware that can respond dynamically to the availability of low emission energy.}

To some extent this is a chicken-egg problem of middleware and application software.
The middleware part is addressed for example in the new SUSFECIT project.
Progress there will further motivate the required adjustments of applications that are suitable for running on dynamically available resources.

\item \textbf{L3: Optimize power usage effectiveness.}

This challenge is well accepted by large computing centers. Newly built computing centers are already targeting PUE values close to 1.0., which is below the global average of 1.6 and close to 1.0, the value for a completely efficient data center. The new CERN Computing Center in Prévessin, for example, has a PUE target of 1.1.~\cite{10.3389/fhpcp.2025.1520207}. 

\item \textbf{L4: Reuse of produced heat.}

Also this challenge seems well accepted by large computing centers. At DESY, for example, several projects are currently underway to improve energy efficiency through the use of waste heat, e.g., by using cooling water from the accelerator systems~\cite{desy-energy-efficiency-projects}.

\item \textbf{L5: Adjust hardware lifetime considering emissions due to procurement and operation.}

The embodied footprint is more and more being looked at for large procurements, but on the one hand information about it may be hard to get and on the other hand financial considerations often take priority over environmental aspects.

\item \textbf{L6: Include the resources needed for continuous IT support into project planning.}

The awareness about the essential role of continuous IT support for data-intense sciences is rising.
Still, it remains a challenge to secure sufficient resources for this task.

\end{itemize}

% ==============================================
% ================================== SECTION TWO
% ==============================================

\section{Sustainability Actions}\label{sustainability-actions}

Sustainability is naturally a very broad topic, covering topics such as sociological impacts and the more commonly talked about greenhouse effects or resource exhaustion (e.g. rare minerals, oil).
For example, the greenhouse effects of diverse gases used in various experiments and detectors, is certainly relevant, but not specifically to the computing and data focused ErUM-Data community.
As such, the  primary focus will be on CO₂ emissions in the context of computing or IT infrastructure related activities.
It should be noted that the water consumption footprint is generally also a major point of interest for computing center operations, but it is of limited relevance, at least for the German community.

\paragraph{CO₂ Emission} can be divided into two major components: Embedded and Operation, which will be explored in detail at a later point.
Beyond that, there are also components that fit into neither of the above, e.g. the footprint of researchers traveling to international conferences. But these aspects are mostly not unique to topics of ErUM-Data, and are best addressed at a broader scope. 

However, the more abstract aspects of working with data, i.e. the analysis of data using software algorithms and tools, are by their nature mostly location invariant.
In consequence, working from another location such as one's home is entirely feasible and has significant potential for reducing tertiary emission sources, e.g. from commutes.
This same location invariance can also be leveraged on the other end, since it makes no difference from the analysis point of view where computations run e.g. in the office or in a (mini) datacenter next to a wind turbine.
These kinds of flexibilities are a major boon of working with data and should be leveraged to its fullest potential.

\paragraph{Embedded Emission} covers the hardware aspects of computation needs.
In the first order this stems from the procurement, i.e. production and construction, but also includes the disposal at the end of life.
It can also be sensible to include (part of) the maintenance footprint, in particular for replacements based on the relevant failure and wear rates.
To accurately account for this, it is necessary to demand corresponding information of the providers, be it hardware manufacturers or building companies.

The construction (and decommission) of a building for a data- or computing-center can in large parts leverage the same sustainability considerations as any other building.

The embedded footprint of the computing hardware itself can best be lowered by increasing the effective longevity.
Of course, this also has implications for the operation emissions, as older hardware is significantly less power efficient.
Here, an increased modularity of the various constituents enables partial upgrades, thus facilitating reuse for at least some parts.
One should carefully examine, to what extent increased heterogeneity and failure rates are tolerable, since there is great  savings potential.
In general, reusing and recycling infrastructure and hardware, and enabling such practices at the end of life/operations, is a major opportunity to decrease embedded emissions.

\paragraph{Operation Emissions} are almost entirely driven by the energy consumption during usage.
This includes the direct consumption by data storage or computing hardware, as well as secondary needs such as lighting, ventilation, and especially cooling.
The total consumption is then scaled by the emissions associated with the energy source, or rather the mix of sources.

Reducing direct consumption is generally difficult, even though many opportunities exist.
One can improve computational efficiency, for example by optimizing algorithms or using more suitable or specialized hardware.
In practice, such gains are often offset by a corresponding increase in demand, so overall consumption does not decrease--often referred to as the Rebound Effect.
Nevertheless, improving hardware and software efficiency remains important and should be incentivized (see \cref{incentivising-action}), for example through user-focused measures discussed in \cref{certifying-software-efficiency}.
Furthermore, the growth in demand can be mitigated by significantly enhancing data reusability, as discussed in \cref{long-term-data-storage}.

In contrast, the cooling footprint can actually be reduced through several approaches.
Waste-heat recovery, which supplies heating to adjacent infrastructure, can lower the overall footprint and should be applied wherever possible.
Locating computing centers in cooler climates can also reduce cooling needs.
The trade-off between this and offsetting increased cooling needs with additional solar or wind power should be carefully evaluated.

Finally, overall operation emissions can be lowered effectively by using greener power sources, although their output is highly variable over time.
This can be addressed either by direct compensation -- e.g., over-commissioning and energy buffers, which however increase embedded emissions -- or by modulating compute throughput to match the availability of low-emission energy, as discussed in detail in~\cref{breathing-computing-centers}.
When selecting a location for a new facility, the local availability of green energy should also be considered, since energy transmission incurs losses, is limited in capacity, and slow to expand.

\paragraph{}
The implementation of any sustainability enhancements requires their inclusion from the very start, not just an afterthought.
As such, it is important to have actionable guidelines about emission/efficiency targets from stakeholder side, providing a well defined and quantified stance on the tradeoff between sustainability and goals and monetary budget.
Overall, it is necessary to thoroughly quantify and measure the footprint through and throughout.

\subsection{Detailed Ideas for Implementations}

This section details a selected few measures, that still have significant potential and have not been investigated in sufficient detail yet.

\subsubsection{Breathing Computing Centers}\label{breathing-computing-centers}

The core idea behind breathing computing centers is to be adaptable to the variability of low-carbon-intensity electricity sources.
Solar energy in particular has a prominent 24-hour availability cycle, which is reminiscent of breathing - it is cyclic but with a fair bit of variation.
In a similar sense, other patterns such as seasonal and weather systems also induce fluctuations of varying regularity.

In most cases, such adaption primarily consists of various forms of energy buffering to bridge the necessary time spans between periods of low carbon intensity in the energy supply.

However, it is also possible to buffer the computation needs, thus enabling modulation of the activity and corresponding energy consumption.
For many applications, the latter is unfeasible due to real-time requirements, but some scientific research workflows can afford to be more accommodating in these aspects. 
In reality, any solution will require some mixture of the two avenues above.

There are many technologies and policies dedicated to buffering energy use across various timescales and each has different strengths and weaknesses.
The three relevant time-scales are: minutes, hours, and days.
The shortest is necessary to facilitate safe emergency shutdowns, e.g. for unexpected loss of the primary power source.
The intermediate range cover phenomena such as day-night cycles or the demand spikes around dinner time.
At the long range these buffers would compensate for extended cloudy or calm weather patterns.

Buffering computation is a far more exotic practice, at least in response to energy availability.
An effective implementation requires active input and corresponding concessions from users of computing resources.
In particular, workloads need to be well classified by the duration of delay they can tolerate.
This will likely require a massive paradigm shift to accommodate such limitations at all necessary levels, from users prioritizing their workloads appropriately to stakeholders tempering their expectations on the timescale of results availability.
The adoption of computation buffering, or workload shifting can be helped by providing incentives to engage with this system and can even go as far as to be a funding prerequisite.
Here, the prospects of success are uniquely high in scientific fields, since many workloads already have long run-times, such that the fluctuations introduced by breathable computing, are likely tolerable and less disruptive than in other fields.

However shifting workloads requires more computing hardware than if the computing resources were right-sized to achieve the same throughput, since the work being done is spread out over a longer period of time with respect to continuous utilization.
While there may be some spare capacity in existing resources due to under-utilization, it is likely not enough to exploit the variability of carbon-intensity in power supply to its full extent, thus necessitating expressively over-provisioned resources.
Simultaneously, such spare idle capacity could be leveraged to satisfy high priority bursts of demand, which comes at the cost of original intent.

This is at odds with the already high cost of computation hardware and makes such operation modes difficult to motivate.
However, it would be entirely reasonable to largely obtain the needed hardware from conventional computing centers when they are retiring their old hardware.
In particular, the comparatively inferior efficiency of outdated hardware is of limited relevancy, since it is operated using predominantly low emission energy anyway.

Fundamentally, computing needs are surprisingly flexible such that they can easily be routed to different locations or facilities depending on overall energy conditions.
With this flexibility one can gain the benefits of otherwise atypical locations, e.g. operating directly near/within a wind farm to avoid power grid costs and bottlenecks.
Also, it is trivial to instantaneously pause and resume (not abort and restart) the running of computations and thus partially modulate the energy consumption of a computing center on the order of seconds, which may be valuable for power grid operators to stabilize their networks.
Additionally, some types of research computations are safely interruptable, e.g. machine learning trainings can use checkpoints, which enables extended suspension and even relocation.

As part of the NHR initiative, there is currently a significant amount of transitions from smaller computing centers, in HEP often called Tier 2 facilities, to fewer larger and well-optimized facilities.
This and similar efforts provide ideal opportunities to implement such breathing computing centers, or demonstrators thereof, while offsetting the majority of the computing hardware cost (monetary and ecological footprint) through reuse of soon-to-be retired hardware.

\subsubsection{Long-term Data Storage}\label{long-term-data-storage}

Experimental data often poses a challenge in balancing its long term storage  with the cost of keeping it.
Generally raw data is extensively processed, often in several discrete stages, until it can finally drive scientific insights.
While this commonly comes with a reduction in volume, this gain is often offset by an increased number of variants of derived datasets, e.g. different revisions of calibrations or different analysis subjects.
Consequently, reducing the footprint of data storage is a complex topic, but it can allow for major savings in overall footprint.

The footprint of keeping derivatives of data can be optimized by consistently quantifying the cost of re-reducing the data to the desired stage.
This cost must include the personpower required to maintain, update, or rebuild all the necessary software and configuration.
The extent of this cost point, may be partially an indicative of how complex and niche the corresponding data is to use.
This can then inform the likelihood of data re-use and thus how sensible archiving it may be in the long run.

While storing data is often cheaper than rerunning the experiment itself, this only applies if the data is actually reusable - in terms of relevance, discoverability, replicability, and readability.
To actually find data to re-use, it must be well curated, especially its metadata, which requires an investment of resources, typically by the data producers.
This also means that the archives themselves need to be either few in number, or better yet -- simultaneously searchable, since finding the (relevant) archive itself precludes discovery.

In a similar vein, it may be preferable to keep only data that is enriched to some degree, especially if the exact details of an enrichment process, such as caveat-heavy calibration procedures, are prone to be lost in time.
Such practice, to only provide the science-ready data in the end, is already fairly common and successful in astronomy.

Additionally, lossy compression can offer massive savings and it may be better to have reduced precision than to throw (portions of) data away.
Beyond this, researchers may be queried to nominate must-keep datasets within their field, to form a broad consensus on generally valuable data to keep.

\subsubsection{Certifying Software Efficiency}\label{certifying-software-efficiency}

Efficient software \emph{that runs at scale} can reduce emissions and make large-scale computing infrastructures sustainable.
Defining efficiency is non-trivial, since measuring energy consumption of software is specific to each program and use case.
One practical approach to do this in the short term would be to only focus on commonly-used software components, thus maximizing the benefit of measuring consumption and improving efficiency.
In the medium term, educating scientists that develop software about ways to minimize resource usage in their software can have a knock-on effect on energy consumption on a larger scale.
On the longest time scale, working towards getting energy certifications for software programs will ensure that energy consumption gets attention in the project planning phase and reduces the carbon footprint of the scientific activity over the lifetime of an experiment
or a research-facility.

\subsection{Incentivizing Action within the Community}\label{incentivising-action}

This section presents some ideas to advance sustainability efforts through various means of binding actions -- those that could be mandated at a global steering level.
There is also an extremely wide array of voluntary methods, by which to promote proactivity on sustainability issues.
Those are detailed in \cref{assisting-awareness}.

The European CO₂ trade already implements a framework to assign a monetary value, and thus impact, to CO₂ emission.
A broader application of this mechanism would effectively require participants, in our case the scientific community, to thoroughly consider all avenues of increasing their sustainability.
Here, it is important to broadly apply these rules, with a few exceptions as possible, to fairly and evenly incentivize actually useful behavior, i.e. prevent circumvention and exploitation of loopholes.
Since a sizable portion of emissions are embedded i.e. in the computing hardware, it would be sensible to apply the need for CO₂ certificates to those components too.
Either way, this would apply an additional strain on research budgets, which can be minimized by either roughly accounting for and reallocating the corresponding increase in CO₂ trade proceeds or by directly allocating a sensible volume of CO₂ certificates.

A simpler means to promote sustainability awareness, would be to mandate comprehensive reporting at broadly visible levels.
For example, it could be reasonable to demand that any publication not only acknowledge its funding sources, but also require a brief breakdown of the footprint of the research.
Such a measure, should primarily by used to increase exposure to sustainability as a topic, and thus enable and invite voluntary competition.
This could simultaneously also increase the value proposition of optimization of intermediaries, e.g. commonly used software libraries, such that researchers could promote their optimization efforts by reporting their work's net-negative footprint -- also referred to as ``handprint''.
Using this information to strictly drive decisions e.g. budget allocations, should only be done at a later stage, once it is well understood and comprehensively formalized.
Otherwise, it would quickly invite ``creative accounting'' whereby the reported value is optimized instead of the intended underlying goal.

\subsection{Teaching \& Training}\label{teaching-training}

In order to strengthen, incentivize, and continue action, it is essential that young scientists in particular are made aware of the issues and given access to tools and skills to take on and drive progress in this area. 

\paragraph{Training Activities for Sustainable Software Development and Use}
Advancing sustainability in the digital transformation in ErUM-Data requires not only technological innovations but also systematic teaching and training practices.
Because software development, data analysis, and high-performance computing are central to ErUM research, sustainability depends critically on the competencies, habits, and values of the researchers who design and use computational tools.
A substantial proportion of software users in the ErUM community are students and early-career scientists.
Thus, targeted training is essential for maximizing both scientific output and environmental impact.

\paragraph{Establishing Sustainable Coding Practices}
Large experimental collaborations already provide coding guidelines to ensure maintainability, readability, and reproducibility of their software.
However, these guidelines often have limited reach or applicability, due to highly differentiated tool usage.
Efficiency considerations are to a large extent already embedded in existing guidelines, yet should be explicitly emphasized to support sustainable software practices. 

These guidelines should not only govern official core software (e.g., reconstruction or data-acquisition frameworks) but also extend to analysis code written by “common”-users.
Checklist-style guidance (e.g. in automated checking tools, which can also be used efficiently by users at the entry level), tailored to the specific experimental ecosystem, can support efficient code design across different user groups. 

Where feasible, collaborations should promote shared software solutions, especially for smaller experiments with limited programming expertise.
This point has been identified as extremely important and requires both the willingness to commit to its implementation and additional resources to do so.

Established community-driven toolkits and frameworks require continued support, maintenance, and associated training programmes to ensure that sustainable software development benefits the entire ErUM-Data community.

Exercising sustainable coding must be accommodated, e.g. in time allocations, and acknowledged as actual scientific merits, as such the actual institutional significance must grow to match the importance of computing in science.

\paragraph{Benefits and Examples of Early Training in Sustainable Software Practices}
A large fraction of emissions associated with data analysis originates from bachelor’s and master’s students, many of whom interact with high-performance computing systems for the first time.
Limited prior experience often results in suboptimal use of computational resources, thereby increasing overall inefficiency.
Such inefficiencies carry multiple consequences: they lead to avoidable CO₂ emissions and can temporarily overload shared computing infrastructures, reducing availability for other users and impeding smooth collaborative workflows.

It must be our core interest to educate and inform students and early career scientists how to use software, and through it hardware, efficiently and what the values are we base our developments on.
We need to ensure not only the development of sustainable software but also the correct use of sustainable software. 

Even introductory, domain-independent training can successfully communicate the limited nature of computational resources.
For example, in machine-learning applications, students often train models on full datasets by default, resulting in unnecessarily long training times and elevated energy consumption.
Introducing the principle that model performance gains saturate with dataset size can significantly reduce computational load: tasks such as hyperparameter optimisation can be performed on suitably reduced subsets before final training on the full dataset.
This approach lowers resource usage and associated CO₂ emissions while maintaining scientific standards.

Another example of useful help for students is an introduction to the usage of job scheduler systems like HTCondor.
When getting in contact with schedulers like this, many students choose one of two ways: run their scripts locally since the set up is easier or just assign a lot of resources so the workload will always run.
The first point defeats the purpose of those schedulers, the second leads to more resources of the cluster being blocked than needed and with that reduced possible performance of the whole cluster.
A single lecture on awareness can lead to the correct usage of those job managing systems and with that improved quality of life not only for the student users, but also for all other users of the cluster.

These examples demonstrate that students can be sensitized to sustainability considerations with comparatively modest training efforts, resulting in both a more efficient computational work environment and reduced emissions.
Such training is itself a sustainable investment, as teaching the fundamentals of sustainability early pays off throughout the whole career and well-prepared students are able to transfer these competencies to colleagues, promoting long-term cultural change within the research community.

For this target group, contemporary instructional approaches must also be considered.
For example, methods such as gamification are currently being explored to facilitate the teaching of automated software testing and to enhance engagement and learning effectiveness.

\paragraph{Developing Teaching Competencies and Communication Skills}
Communicating the motivation behind sustainability efforts is crucial.
Young scientists in particular (but also established researchers) need to understand why efficient computing matters in terms of environmental impact.
Embedding this context early in technical training supports acceptance and long-term adoption of sustainable practices.

Technical expertise alone is insufficient for effective teaching of sustainable computing practices.
Instructors must also possess strong communication and pedagogical skills.
The consensus is that expertise in just one of these areas is not enough to convey the importance of sustainability issues.
The goal must be to ensure that technical training takes place after all training participants understand the urgency of sustainability issues.
Training programmes for tutors and lecturers should therefore include dedicated workshops on communication, didactics, and student engagement.
It is not only important that professors, speakers, and trainers possess strong communication skills, but also that tutors receive such training, as they often maintain a particularly close and frequent interaction with the learners. 

The ErUM-Data community already offers a comprehensive portfolio of technical training.
To complement this, workshops on communication skills should be implemented, with participation required for all speakers and tutors prior to their engagement in student instruction.
Existing mechanisms for evaluating teaching quality should be used and reinforced to ensure continuous improvement. 

\paragraph{New Concepts for Support and Training}
As increasingly sophisticated simulation and data-analysis methods are developed, it is vital that researchers are capable of employing these tools effectively.
Sustainable software gains real impact only when users understand how to operate it efficiently.
The existing technical training offerings for the ErUM-Data community, including schools and workshops organized by the ErUM-Data Hub, have been well received. These established programs can be further complemented by the introduction of new training concepts.

While schools and workshops provide structured learning opportunities, many inefficiencies in software development stem from a lack of onboarding and individualized support.
Middle-term “code consulting” or mentoring programmes can address these gaps.
Such programmes offer one-on-one guidance on setting up efficient analyses, optimizing Monte Carlo simulations, or improving the sustainability of existing code.
Regular virtual meetings enable mentors to provide feedback, while optional group sessions allow mentees to present progress and learn from one another.

Over the long term, a searchable registry of experts within the ErUM-Data community could provide an accessible entry point for newcomers seeking topic-specific support.
This would reduce common inefficiencies and promote sustainable research practices across the community.

In addition, establishing a network of trained student ambassadors can facilitate the continuous dissemination of knowledge on sustainable computing within all ErUM communities.
Ambassadors can share best practices, highlight relevant training materials, and act as contact points within their community.
However, such a network requires centralised organisation, active maintenance, regular check-ins, and clear communication channels to remain effective.

\paragraph{Driving Knowledge Transfer Beyond the ErUM-Data Community}
The ErUM-Data community strives to play a leading role in developing sustainable computing strategies and technologies.
To maximize societal benefit, this knowledge must be disseminated beyond disciplinary limits. Therefore, the results should not only be published and made available, but programs should also be set up to proactively educate individuals responsible in other areas of research as well as the industry.

To achieve this, specific target groups with similar needs and capacities must be identified, and room must be created to bring together speakers from the ErUM-Data community who possess expertise relevant to these groups.
This approach would not only facilitate the dissemination of knowledge and technology, thereby increasing their overall impact, but also strengthen the position of the ErUM-Data community as a key contributor to societal progress.

\subsection{Steering \& Funding}\label{steering-funding}

In the course of discussing challenges, ongoing and potential new measures, the conversation repeatedly returned to questions of funding. In particular, participants raised the question of the extent to which ErUM-Data scientists are expected to address sustainability explicitly within their projects. For example, should there be a recommended benchmark for the proportion of personnel time and/or financial resources that ought to be allocated to sustainability-related measures? The potential impact of such requirements on new projects was assessed as likely to be high. 

Moreover, the quantifiability of sustainability measures is closely linked to the availability and structuring of financial resources. As such, it is important that funding bodies provide actionable guidelines on emission and efficiency targets, including a well-defined and quantified position on the trade-off between sustainability objectives, scientific goals, and budgetary constraints.

Within the ErUM-Data context, no dedicated resources explicitly earmarked for stand-alone sustainability measures appear to be available. However, the implementation of long-term and strategically planned sustainability activities was considered impossible without such targeted funding. Specific funds could be made available for sustainability activities and, ideally, these funds should be accessible independently of ErUM-Data research projects.

In addition, positive incentives to implement sustainability measures from the outset of project planning do not yet appear to be established in the current funding landscape. Financial incentives could be created to promote resource-efficient (e.g. CO₂-saving) research. Projects with particularly strong sustainability concepts could, for instance, be rewarded with additional funding or extended project durations.

% =================================================
% ================================== SECTION THREE
% =================================================

\section{Ethics}\label{ethics}

The ethics section of the workshop explored what ethics means in the context of ErUM-Data and how it shapes the our research practices. We started from the understanding that ErUM-Data researchers are tasked with conducting accurate and socially relevant research and with mentoring early-career scientists. In doing so, researchers bear full responsibility for their work, for the gains in knowledge, and for what they publish and in what form. Research must be conducted with integrity in accordance with good scientific practice. Results must be reproducible.

Within this framework, we discussed the growing use of artificial intelligence (AI) in the ErUM sciences, highlighting that its development and application present not only undoubtedly great opportunities, but also distinct ethical challenges that must be carefully considered.

\subsection{Transformative Potential of AI}\label{transformative-potential-of-ai}

To begin with, we note the fundamental question posed by the German Ethics Council for an ethical emulation: Does artificial intelligence enhance or diminish human authorship and the conditions required for responsible action? Building on this guiding question, several aspects emerged as particularly relevant and insightful for ErUM research: 

\paragraph{Autonomy} Refers to the possible restriction of autonomy when tasks are delegated to artificial intelligence to such an extent that it results in a loss of skills of the
scientists.

\paragraph{Bias} AI systems are only as objective as the data they are trained on. It is widely recognized that the training of AI tools should not contain signs of bias, which can be examined and prevented using a so-called fairness matrix. This area also includes the issue of plagiarism that may have been incorporated into the training of the AI tool.

\paragraph{Epistemic Aspect} Particular emphasis is placed on the importance of documenting in a traceable manner the pathway by which information was acquired or knowledge was generated, and the degree of certainty with which the correctness of the results can be established.

\paragraph{Measurability of Information Gain and Environmental Costs} Overall, this aspect is particularly challenging and verges on the impossible. While direct quantification is often difficult or infeasible, mathematical frameworks exist that allow  to quantify the relevance of results, their value, or even just the amount of newly acquired information. Such evaluations must account for the strategic importance and sequencing of projects. Finally, the use of resources and the associated CO₂e emissions required for gaining knowledge must be related to and evaluated against the value of the knowledge gained.

\paragraph{Mentoring} The development of AI tools for educational and continuing-education purposes is attractive. AI programming assistants and chatbots tailored specifically to the
ErUM sciences likewise offer an appealing prospect. However, it should not be over-
looked that education in general (subject-matter knowledge, personality development,
a culture of dealing with errors, etc.) is also a social process between early-career re-
searchers and mentors, something from which AI tools are presumably still far removed.

Particular attention was paid to the aspects of transparency and explainability of AI and how this relates to collective and individual responsibility in its development and use. These aspects are explored in more detail in the following chapters.

\subsection{Transparency and Explainability}\label{transparency-and-explainability}

In the context of ethical considerations surrounding software provision and usage, transparency and explainability constitute key challenges of notable complexity. Many AI systems operate as black boxes, rendering their decision-making processes difficult to interpret. Closely related is the issue of result explainability, which requires the specification of an appropriate level of granularity relative to the intended audience. This granularity must be calibrated such that the scientific community is provided with full methodological and epistemic transparency, while the broader public and interested non-experts can access and comprehend the resulting knowledge in an accessible and intelligible form.

Communication within a scientific community is facilitated by shared terminology. When addressing civil society, the use of appropriate and accessible language becomes critical, encompassing both visual representation and the communication of societal implications of scientific findings. In fields such as MINT research and physics, which are often perceived as black boxes, enhancing comprehensibility is essential for rebuilding public trust. In this context, citizen science initiatives can provide valuable support, as demonstrated in areas such as (astro)particle physics.

A recurring concern is bias. In a future in which AI systems become dominant or at least widely deployed, increasing the accessibility of training datasets is essential to enable systematic scrutiny and the identification of unfair biases. Clear contact information should be published with the dataset to ensure that identified biases can be easily addressed. While this alone is not sufficient to render AI-based research or commercial applications fully trustworthy, it represents an important and necessary step in that direction.

Addressing the inadvertent introduction of unfair biases into computational models requires a multifaceted approach. Key measures include (1) the identification and removal of morally irrelevant attributes from training data and (2) the reduction of model opacity through the development and application of explainability methods. A positive side effect of attempting to remove unfair bias would be a model that is describable at different levels of granularity, which can be used to reach different target audiences and the general public, thus supporting our science case in future stakeholder discussions.

Fair and well-curated input datasets were identified as essential for fostering trust in AI systems and for preventing discriminatory outcomes arising from their use. To set a precedent, the ErUM community should proactively develop and adopt best practices for the release of input datasets used in machine learning applications. This might require developing new compression and storage formats to limit storage requirements and the associated consumption of resources. Efficient mechanisms to ensure that input training sets are not stored multiple times, and are easily accessible including the necessary documentation are important as well. The release of training datasets may occur with a defined time delay to allow for initial proprietary analyses, provided that the time and conditions of data release are clearly communicated. Establishing these practices as standard, accompanied by clear and transparent external communication, may encourage other scientific disciplines to adopt similar approaches – particularly in domains where input datasets are more susceptible to unfair biases, such as sexism or racism. 

Finally, transparency also extends to issues of authorship. In the coming years, it may become increasingly difficult to verify the contributions of individuals or systems to a given piece of research. This raises critical questions regarding the extent to which trust can be relied upon, or whether full verification may ultimately be unattainable. This issue is closely linked to questions of responsibility in the development and use of AI, which will be explored in greater detail in the following section.

\subsection{Attribution of Responsibility}\label{attribution-of-responsibility}

In the discussion about the responsibility that institutions and individuals (in our case, ErUM-Data scientists) assume in the development and application of AI, one statement from a presentation received widespread agreement and subsequently served as a guiding principle: “Machines cannot relieve humans of their responsibility.” It was generally acknowledged that researchers remain accountable for the outcomes of AI systems and must retain the capacity to verify results. Furthermore, the discussions yielded that the user of the AI is and stays responsible for checking the outcome of the respective AI prompt.
The following points were discussed in this context:

\paragraph{Accountability in Software Usage} Users bear responsibility and accountability for the scientific outcomes and results produced using software tools. Consequently, it is essential to document in every publication all software employed, including full configuration parameters and, where applicable, the use of AI tools. Configurations, prompts, and discussions that contributed to the outputs generated by AI systems constitute part of the research output and must be incorporated into research data management processes.

Given that reproducibility is central to accountability, users are expected to utilize software that adheres to FAIR principles and to clearly document how the software was applied. ErUM users should always strive for the usage of publicly available software tools. Software developers are accountable for providing comprehensive and accessible documentation of all software features. Users should be permitted to correct or improve identified deficiencies in the software. These improvements also fall under the requirements of documentation.

\paragraph{Collective Responsibility} In contemporary collaborative research, responsibility for results is frequently distributed across multiple contributors. Consequently, accountability for the outcomes of team efforts is shared, while each individual remains responsible for their specific contributions. In this context, transparency regarding the AI methods employed is particularly critical to ensure clarity and trustworthiness of the collective work.

\paragraph{Accountability in the Use of AI/ML Algorithms in Research} The use of AI and ML algorithms, including the inputs/training datasets, methods, and architectures, must be documented with the same scientific rigor applied to other research methodologies. As AI becomes increasingly prevalent, enhancing transparency may require the release of training datasets, potentially in compressed formats to reduce storage demands, both within the ErUM community and more broadly across other scientific disciplines and collaborative initiatives (see 3.1). Responsibility for the application of AI/ML rests with the researcher, or, in industrial contexts, with a designated individual capable of verifying the applied algorithms (e.g., a product owner). Accountability, including scientific or corporate reputation and legal liability, resides with this responsible person to ensure that AI/ML tools are employed correctly and securely. Correspondingly, successful outcomes are credited to the individual or organization that designed, implemented, and authorized the use of AI/ML, rather than to the AI system itself.

\paragraph{Accountability in the Use of AI for Documentation and Scientific Publications} The use of generative AI, including LLMs, should be regarded as an advanced “assistive tool” for scientific writing. However, such tools should never be relied upon to produce content in its entirety, nor can they assume responsibility for the resulting text. Accountability remains with the researcher, who must ensure that any AI-assisted content is accurate, scientifically sound, and free from unintended plagiarism.
LLMs cannot be listed as authors in scientific publications; the human author is responsible for verifying that all text, figures, code, or other content produced with AI support meets established scientific standards. Appropriate tools, such as plagiarism detection software, should be made available to researchers by their institutions. 

\paragraph{Use of AI in Bachelor and Master Theses} A key question concerns the appropriate use of AI in student theses. Unlike researchers, who typically possess advanced knowledge of their subject and can leverage AI to optimize certain aspects of their work, students are still in the process of acquiring foundational understanding. Therefore, students may need to complete tasks without relying entirely on AI, to fully learn essential processes and develop a deeper comprehension of their topic. Students should be able to critically assess their own work, ensuring that AI-generated contributions are properly understood, evaluated, and integrated into their learning process.

If an accountability-based approach is adopted for scientific publications, the same principles should apply to student theses, as these are intended to train students in scientific work. Accordingly, a thesis should not be entirely AI-generated, either in terms of text or research work, and students must assume full responsibility for the accuracy and scientific integrity of their work. To support this, supervisors and students should be provided with appropriate tools by their institutions, including AI detection software to evaluate the extent of AI-generated text, as well as plagiarism detection tools. The supervisor retains full responsibility for the assessment. AI tools cannot assume accountability for grading decisions.

\paragraph{Responsibility for Sustainability} All software tools, including AI, must be used responsibly, with careful consideration of resource efficiency and associated CO₂-equivalent emissions. Users are accountable for minimizing these impacts, which requires prior planning, reasonableness checks, and small-scale testing before large-scale computations.

Ensuring accountability for CO₂-equivalent emissions requires the implementation of systems that promote responsible use of software and environmental resources. Such systems should incorporate concrete methods to facilitate the measurement of consumed resources and their corresponding impacts (e.g. financially). 

One illustrative scenario could involve attributing a personal share of the global atmosphere to each individual. If the respective part of the atmosphere is polluted with CO₂, the owner could seek compensation from the national government, thereby creating an incentive for the government to recover costs from the party responsible for the emissions.

\subsection{Further Implications}\label{future-perspective}

In addition to the previously discussed aspects, further considerations have emerged regarding the use of AI within ErUM-Data.

\paragraph{Risk of Deskilling} The risk of deskilling is a significant concern for ErUM sciences and must be actively addressed. Researchers must retain the skills necessary to assume full responsibility for their results. Key skills that should continue to be promoted among students in the era of AI include:
\begin{itemize}
    \item Critical thinking
    \item Experimental design
    \item Communication of experimental findings and data-driven conclusions
\end{itemize}
It is essential to assess in detail which skills have been central to scientific training and which additional competencies are now required to ensure both scientific rigor and accountability in an AI-supported research environment.

\paragraph{Critical Thinking as a Cornerstone} The importance of critical thinking is increasing, particularly as Large Language Models (LLMs) are being employed in code development within scientific software frameworks with growing prevalence. By critically evaluating the results of LLM queries, scientists can verify the correctness of the generated code while simultaneously deepening their understanding of the underlying research and the impact of these outputs on ongoing scientific work. 

The dramatic increase in available data has correspondingly heightened demands on its processing. While AI can support the summarization of key findings, objectives and tasks must be clearly defined to avoid introducing bias. Essential skills in this context include abstraction, precision, and contextualization, complemented by the ability to critically reflect on one’s own analyses and outcomes.

To establish these foundational skills, the systematic teaching of critical thinking throughout physics curricula is essential. This enables early-career scientists to develop a sophisticated level of critical reasoning. Effective implementation of such teaching requires not only sufficient instructional resources, but also the training of instructors to communicate and cultivate critical thinking skills effectively.

\paragraph{Strong Foundation for All Students} In the first term of any course, students should participate in an AI-focused workshop to ensure that all learners, from the earliest stages of their education, receive at least basic guidance on the use of AI, expressly including ethical considerations.
This introductory session does not need to be lengthy but should address critical thinking in the context of AI, highlighting both potential benefits and risks. Such a workshop can be made mandatory at the undergraduate level or adapted for individual faculties. It may also be offered at the postgraduate or PhD level. Importantly, these AI-usage skills are transferable across disciplines and research contexts.

\section{Assisting Awareness}\label{assisting-awareness}

At the conclusion of the workshop, there was a strong consensus regarding the necessity to engage actively in two key areas:
\begin{itemize}
    \item Driving long-term and strategic initiatives: It is imperative that sustainability standards are integrated from the outset in all projects, regardless of their scale, and prioritized accordingly.
    \item Implementing short-term, actionable measures: Additionally, there is a need for immediate actions that can be executed without requiring additional funding. This includes particularly communication strategies targeted at diverse audiences. We must enhance our efforts to effectively convey our work and sustainability initiatives not only to the general public but also to foster discussions within our community by integrating aspects of sustainability and ethics at every opportunity.
\end{itemize}

\subsection{Need of Common Values and Visions}\label{common-visions}

The discussion on taking action was framed around a central question: Why do many individuals refrain from engaging in climate change mitigation, even though scientific data are widely available and consistently point in the same direction? Our working assumption is that, while many people agree that reducing CO₂ emissions in science is essential, various factors hinder the implementation of concrete measures. These include:
\begin{itemize}
    \item a lack of information on how they can effectively contribute,
    \item a lack of awareness of their own individual contribution to the problem,
    \item the perception that individual action is ineffective and the feeling of being left alone,
    \item low motivation due to missing or insufficient incentives, 
    \item purely opportunistic behavior, 
    \item the perception that saving energy is associated with additional, unpleasant work, and
    \item skepticism toward or rejection of scientific evidence.
\end{itemize}
 
This led to the conclusion that a shared understanding of the urgency of climate action is a prerequisite for developing common visions and values within the community. The central challenge is to ensure that CO₂-efficient approaches are internalised as an integral component of scientific practice. This, in turn, requires the creation of appropriate incentive structures. A purely top-down approach should be avoided, as it is likely to result in indifference or, in the worst case, be counterproductive. Instead, efforts to reduce CO₂ emissions should ideally be driven by researchers’ intrinsic motivation; however, we assume that such intrinsic motivation will rarely emerge in the absence of external incentives.

The key question, therefore, is how to foster this internal willingness to adopt energy-efficient practices. The current situation resembles a stalemate: individual researchers often consider themselves too insignificant to have a meaningful impact and therefore shift responsibility to higher-level actors such as governments, professional societies (e.g. the DPG), funding agencies, or similar institutions. At the same time, these institutions frequently frame the problem as one that should be solved primarily through technical means. Yet technical solutions, such as increased software efficiency, tend to lead to the production of even larger data sets, since the prevailing scientific incentive structure rewards producing the “best” possible paper, which is often associated with maximal data volume and compute time.

\subsection{From Awareness to Action}\label{From Awareness to Action}

Where awareness and willingness are already present, the central challenge is to enable individuals to translate this willingness into concrete action. The question is how researchers can be supported in moving from general intention to the practical implementation of sustainable behaviours. Potential approaches identified include:
\paragraph{Engaging Science Communication Formats and Teaching Innovations} Popular-science formats (e.g. short videos or other audio-visual materials that create a lasting memory effect) could complement or partially replace traditional teaching methods. At the same time, efficient use of computational resources should become a standard component of curricula, for example in computational physics courses. Corresponding teaching materials could be made available on a dedicated website.
\paragraph{Provision of Tools to Quantify Individual CO₂ Emissions} Freely accessible, easy-to-use tools should be developed that provide researchers with immediate feedback on the CO₂ emissions associated with their computational work (ideally via a “one-click” result). Such tools could be jointly developed by technical staff and physicists and made available to the wider community. 
\paragraph{Dedicated Formats and Recognition at Conferences} Major national and international conferences should include dedicated sessions on sustainable research practices. In addition, a sustainability prize, for example for the most resource-efficient research project, could be established. 
\paragraph{Contextualisation of CO₂ Metrics} Reported CO₂ values should be translated into intuitive real-world equivalents (e.g. annual per-capita emissions of specific countries, or the emissions associated with certain flight distances) to make their implications more tangible.
\paragraph{Public Disclosure of the CO₂ Footprint} The CO₂ equivalents required to generate scientific results could be reported systematically, for example as part of each publication or scientific presentation. Similarly, third-party funded projects could be asked to include estimated CO₂ emissions in mid-term and final reports. Projects with exceptionally low emissions might be rewarded, for instance through the option of extending the project’s duration.
\paragraph{Role Modelling and Positive Incentives} Direct coercive measures should be avoided. Instead, prominent scientists could lead by example so that sustainable practices gradually become a community norm. In general, positive incentives are preferable to additional bureaucratic accounting requirements.
\paragraph{Highlighting Efficiency as a Scientific Advantage} Efficient codes and usage practices can provide a competitive scientific advantage, as they enable more sustainable research within shorter time frames. A CO₂ quota per researcher or project was discussed as a way to limit rebound effects: remaining “emission budget” could be used for further research, potentially making efficient researchers attractive collaboration partners.
\paragraph{Design of “Win–Win” Workflows} Easy-to-follow guidelines should be developed that simplify scientific workflows while automatically integrating sustainability considerations, thereby creating situations where the most convenient option is also the most sustainable one.
\paragraph{Establishing Sustainability Experts and Multipliers} Sustainability experts could be invited regularly as speakers at conferences, allowing others to benefit from their experience. Early-career researchers with a focus on sustainability can act as multipliers both within and beyond the scientific community.
\paragraph{Institutionalisation through Certification and Training} Easy applicable sustainability certification schemes for laboratories could be introduced. In addition, sustainable research practices should be systematically taught to all new students as part of their transferable skills or “soft skills” training.

\subsection{Engaging Beyond Information: Motivating Audiences}\label{Engaging Beyond Information: Motivating Audiences}

One statement from the discussion particularly resonated within the workshop group:
“The problem and many of the solutions to the climate crisis are well known. The key question is how we can achieve broader adoption of these solutions. Simply communicating the facts and expecting a rational response does not work. We have to acknowledge the diverse values of the people we want to reach, choose appropriate framing, and speak as members of the addressed communities rather than as external advisers. We should begin within our own research groups, experiments, and communities.”

We are convinced that rethinking both the content and the form of our communication is the key to conveying the urgency of sustainability measures to target audiences inside and outside the scientific community. This requires us to reflect on which audiences we intend to address and how to effectively engage in dialogue with them. It is not sufficient to prepare our messages and goals once and for all and simply “send them out into the world.”

Two complementary and indispensable approaches emerged. First, we should seize every opportunity to enable rapid, low‑bureaucratic progress. This includes actively communicating about sustainability in our own research groups, pushing the topic forward, and creating dedicated space and time for it. Individuals in decision-making positions need to be aware of their responsibility to prioritise such discussions in their groups, and to act on this responsibility without delay. The same applies to e.g. conference presentations and to the measures outlined in Section 4.2. The guiding principle should be: it is not enough to reduce our footprint, we need to also improve our positive impact. During the workshop, this concept was summarized under the term “handprint.” In contrast to quantifying environmental damage, the handprint assesses positive contributions, including improvements in resource efficiency, reductions in emissions, and the effectiveness of communication measures. 

One concrete outcome of this line of thought is the establishment of a “Sustainability Action Group,” which will focus on immediately implementable measures such as preparing a slide deck for dissemination within the ErUM-Data community to support group leaders in identifying entry points in their own groups, and recording a podcast episode on the topics discussed during the workshop. This measure has been implemented since the workshop. 

Second, we need to move beyond our existing communication routines and build strategic campaigns that not only transmit information but also deeply engage specific target groups, motivate them, and enable concrete action. The effectiveness of such measures depends on clear target-group orientation (sufficient granularity), planning reliability, and adequate resources (personnel and financial). Only under these conditions can we create “paved roads” for other ErUM-Data scientists and for the societal actors within our sphere of influence.

\section{Conclusions}\label{conclusions}

This workshop report reviewed the progress made since the 2023 Workshop “Sustainability in the Digital Transformation of Basic Research on Universe and Matter” and assessed the implementation of sustainability measures in ErUM-Data across short-, medium-, and long-term time scales. Significant advances have been achieved in raising awareness, improving transparency of energy use, adopting optimized software and workflows, and integrating sustainability considerations into research planning. Medium-term efforts have strengthened data and software FAIRness, resource-efficient data handling, workflow management, and the responsible use of AI, while highlighting persistent challenges in monitoring, long-term maintenance, and training. Long-term goals such as dynamic use of renewable energy, optimized infrastructure, heat reuse, and sustainable hardware lifecycles remain difficult but are increasingly embedded in strategic planning. Overall, the review demonstrates meaningful progress driven by community engagement and technological innovation, while underscoring the need for coordination, systematic monitoring, and institutional support to fully embed resource-aware practices in ErUM-Data.

The discussion on sustainable actions at the 2025 workshop “Shaping the Digital Future of ErUM Research: Sustainability and Ethics” outlined a comprehensive framework for sustainability actions in data-intensive ErUM research, focusing primarily on reducing CO₂ emissions from computing and IT infrastructure. The ideas distinguished between embedded and operational emissions and highlighted strategies such as extending the lifetime of hardware, improving energy-efficient software and workflows, improving data reuse, and integrating sustainability considerations from the initial planning stages. Innovative concepts such as “breathing” computing centers, optimized long-term data storage, and software efficiency certification were presented as promising approaches to align computational activity with low-emission energy availability. In addition, the importance of community-wide incentives, transparent reporting, and policy mechanisms to encourage responsible behavior were emphasized during the workshop. Substantial emission reductions can be achieved through a combination of technological innovation, structural adaptation, and coordinated governance.

In addition, it was emphasized that long-term sustainability in ErUM-Data research depends not only on technical solutions, but also on systematic education, community support, and appropriate funding structures. Targeted training for students and early-career researchers, the promotion of sustainable coding practices, and the development of strong teaching and communication competencies were identified as key drivers for long-term cultural change. Complementary mentoring, expert networks, and knowledge transfer beyond the community further strengthen the impact of these efforts. The central role of steering and funding in enabling sustainable practices was highlighted, calling for clear benchmarks, dedicated resources, and positive incentives that integrate sustainability into project planning from the outset. Together, these measures underscore the need for coordinated investment in people, skills, and governance to embed sustainability permanently in ErUM-Data research.

The second part of the workshop examined the ethical dimensions of data-intensive and AI-supported research in the ErUM-Data context, emphasizing integrity, reproducibility, and human responsibility as central principles. Both, the transformative potential and the risks of AI, were highlighted, including issues of autonomy, bias, transparency, environmental impact, and deskilling. The third chapter of this workshop report stressed the necessity of explainable and trustworthy systems, well-curated and accessible datasets, and clear standards for authorship and documentation. It reaffirmed that accountability for scientific outcomes remains with human researchers, whether in research, education, or publication. By underscoring the importance of critical thinking, mentoring, and ethical training from early stages of education onward, the discussions demonstrated that responsible AI use is essential for maintaining scientific integrity, public trust, and sustainable research practices.

A central conclusion drawn from the workshop was that sustainable transformation within the ErUM-Data community relies on targeted awareness-building, strategic communication, and appropriate motivation strategies. It was agreed not only that raising awareness must be a priority, but also that the approach to this must be adapted to the new requirements. The fourth report chapter highlighted the importance of developing shared values, aligning incentives, and overcoming barriers that hinder individual engagement, such as perceived inefficacy and lack of guidance. Moving from awareness to action requires practical tools, visible recognition, role models, and “win–win” workflows that integrate sustainability into everyday research practices. Effective engagement must go beyond information transfer, relying on audience-specific communication and dialogue. Promoting both short-term actions and long-term cultural change, the workshop emphasized the crucial importance of communication and motivation in integrating sustainability into scientific practice.

\pagebreak
\bibliographystyle{unsrturl}
\bibliography{references}

\end{document}